\documentstyle[12pt,twoside,fleqn,espcrc1]{article}
\input epsf

\newcommand{\AmS}{{\protect\the\textfont2
  A\kern-.1667em\lower.5ex\hbox{M}\kern-.125emS}}

\title{
Emission of single photons, hadrons, and dileptons
in $Pb+Pb$ collisions at CERN SPS and quark hadron phase transition}

\author{Dinesh Kumar Srivastava,\address{Variable Energy Cyclotron
Centre,\\ 1/AF Bidhan Nagar, Kolkata 700 064, India}
Bikash Sinha,\address{Variable Energy Cyclotron Centre,\\1/AF Bidhan
Nagar, Kolkata 700 064, India\\
and\\
Saha Institute of Nuclear Physics, \\ 1/AF Bidhan Nagar, Kolkata
700 064, India\\}
Ioulia Kvasnikova, and Charles Gale\address{
Physics Department, McGill University, H3A 2T8 Montr\'eal, Canada}}

\begin{document}
\maketitle

\begin{abstract}

The production of single photons in $Pb+Pb$ collisions
at the CERN SPS as measured by the WA98 experiment is analysed.
A quark gluon plasma is assumed to be formed initially, which expands,
cools, hadronizes, and undergoes  freeze-out.
A rich hadronic equation of state is used and the transverse expansion of the
interacting system is taken into account. The recent estimates of photon
production in quark-matter (at two loop level) along with the dominant
reactions in the hadronic matter leading to photons
are used. About half of the radiated photons  are seen to have a thermal
origin. The same treatment and the initial conditions provide a very good
 description to hadronic
spectra measured by several groups and the intermediate mass
dileptons measured by the NA50 experiment, lending a strong support
to the conclusion that quark gluon plasma has been formed
in these collisions.
\end{abstract}

\section*{}

 It has been
recognised for a long time  that electromagnetic radiations from
relativistic heavy ion collisions would be
a definitive signature of the
formation of a hot and dense plasma of quarks and gluons, consequent to a
quark-hadron phase transition.
Once other signs of the quark-hadron transition, e.g.,
an enhanced production of strangeness, a suppression of $J/\psi$
production,  radiation of dileptons,
etc., started to emerge, it was imperative that
the more direct, yet much more difficult to isolate, signature
of the hot and dense quark-gluon plasma, the single photons
were identified. The WA98 experiment~\cite{wa98} has now reported
observation of single photons in central $Pb+Pb$ collisions at the
CERN SPS.

In the present work we show that these data, along with the
hadronic spectra measured by several groups~\cite{na44,na49,wa98pi}
and the intermediate
mass dileptons  measured by the NA50~\cite{na50} experiment are
very well described if we assume that a quark-gluon plasma was
formed in the collision.

\setcounter{figure}{0}
\begin{figure}[htb]
\begin{minipage}[t]{80mm}
\epsfxsize=3.25in
\epsfbox{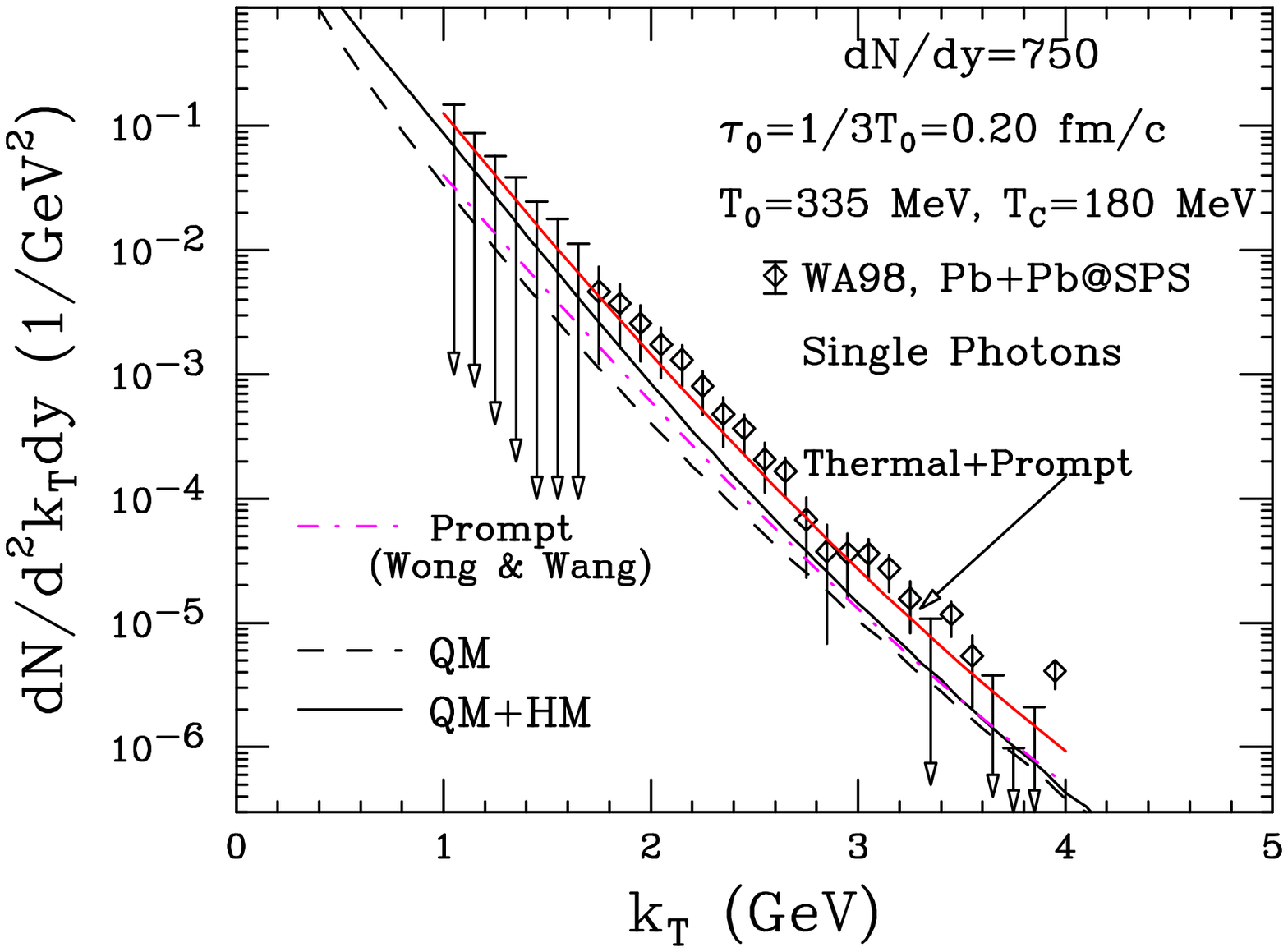}
\label{wa98}
\end{minipage}
\hspace{0.04\linewidth}
\begin{minipage}[b]{0.41\linewidth}
\caption{ 
 Single photon production in $Pb+Pb$ collision at the CERN SPS.
 QM stands for radiations from the
quark matter in the QGP phase and the mixed phase, HM denotes
the radiation from the hadronic matter in the mixed phase and the
hadronic phase. Prompt photons are estimated
using NLO pQCD with the inclusion of 
intrinsic \protect$k_T$ of partons (Wong and Wang~\protect\cite{ktcr}).
}
\end{minipage}
\begin{minipage}[t]{80mm}
\epsfxsize=3.25in
\epsfbox{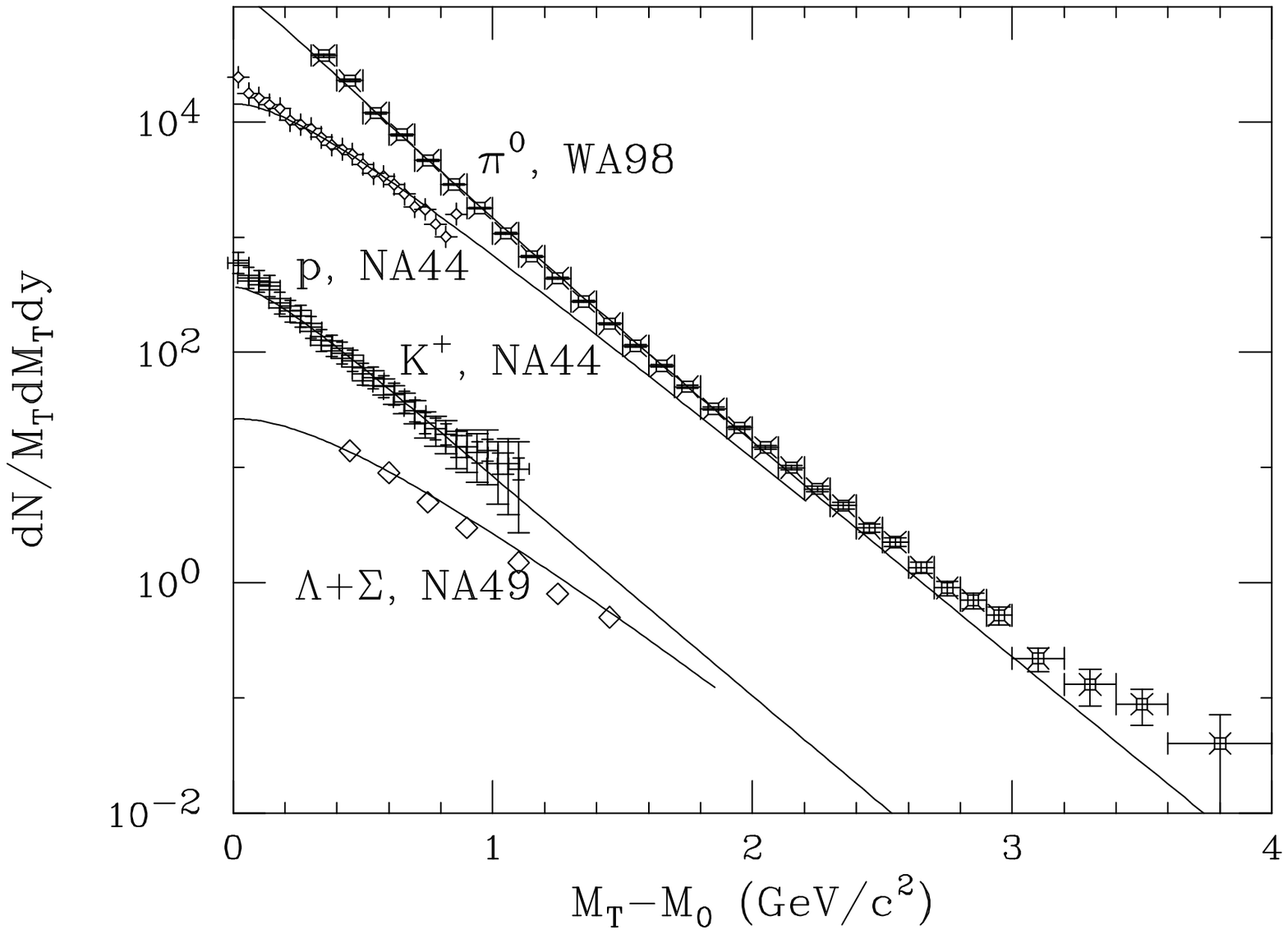}
\label{all}
\end{minipage}
\hspace{0.04\linewidth}
\begin{minipage}[b]{0.41\linewidth}
\caption{
Transverse momentum spectra of neutral pions, protons, kaons
and \protect$\Sigma+\Lambda$ in central
collisions of lead nuclei at CERN SPS. The initial conditions
used for all the figures are identical.
}
\end{minipage}
\end{figure}
\vspace{-2mm}
We incorporate two important improvements in our analysis.
 Firstly, the hadronic equation of state is generalized to
include all of the hadrons~\cite{crs} in the particle data book. 
Secondly,  we use the rate of single photon production from the
quark matter to the order of two-loops reported recently
by Aurenche et al~\cite{pat}. 
 {\em A recalculation of these
rates have been reported recently~\cite{MS} and it has been claimed that the
result of Aurenche et al. is too high by a numerical factor of 4.
We use the corrected rates in what follows. The results presented
during the conference used the former rates of Aurenche et al. (This affects
only the Fig.~1.)}
 For analysis of the dilepton
production we use the exhaustive rate calculations of Ref.~\cite{ioul}.

We assume that a chemically and thermally equilibrated quark-gluon
plasma is produced in such  these collisions at the time $\tau_0$,
 and use the isentropy condition~\cite{bj};
\begin{equation}
\frac{2\pi^4}{45\zeta(3)}\,\frac{1}{A_T}\frac{dN}{dy}=4 a
T_0^3\tau_0
\label{T0}
\end{equation}
to estimate the initial temperature, where $A_T$ is the
transverse area.
We have
taken the  average particle rapidity density as 750 for the 10\% most
 central $Pb+Pb$ collisions at the CERN SPS energy
as measured in the experiment. We use a mass number of 190 
to account for non-zero impact parameter based on estimates
of number of participants.
\setcounter{figure}{2}
\begin{figure}[htb]
\begin{minipage}[t]{80mm}
\epsfxsize=3.25in
\epsfbox{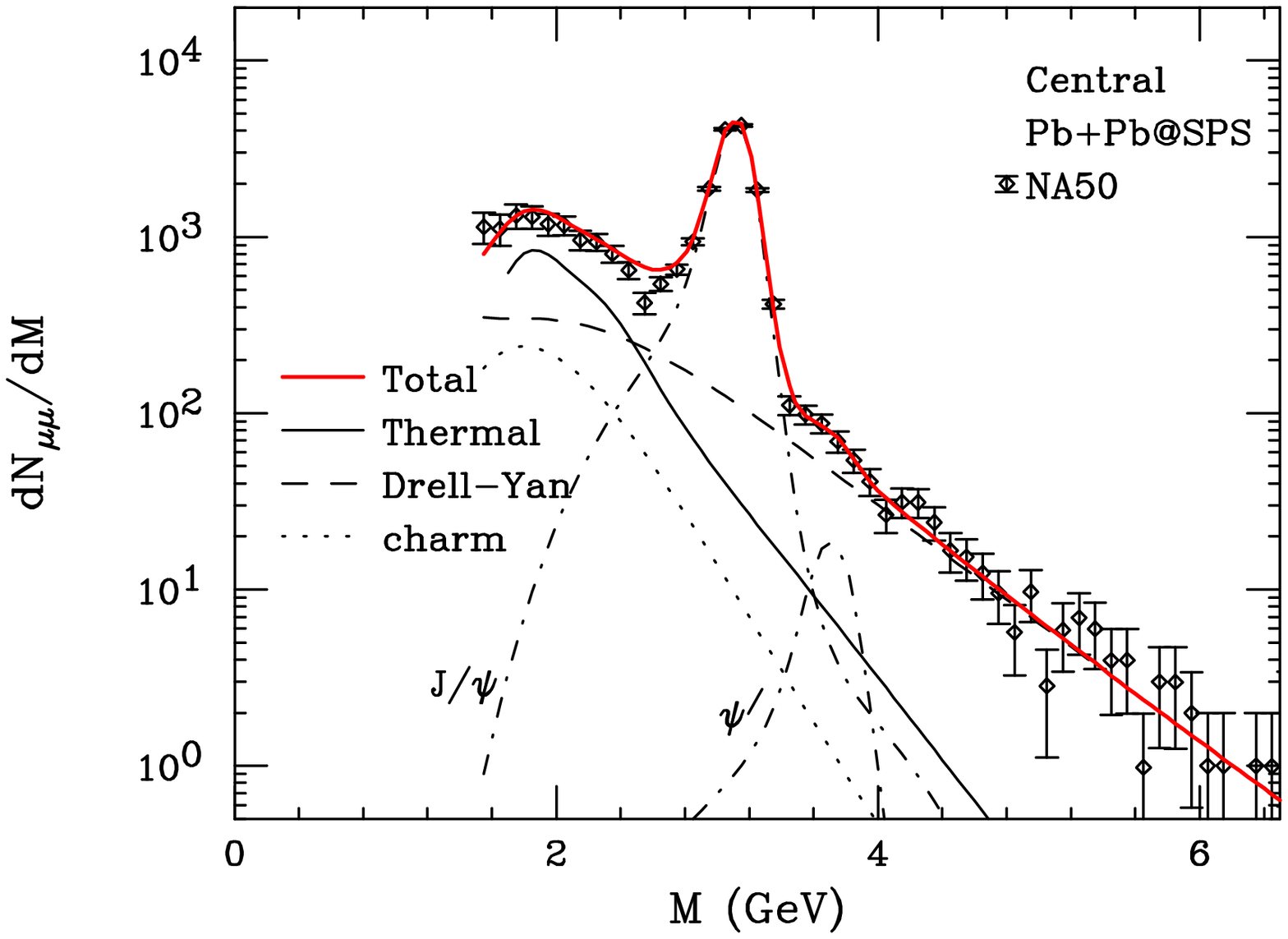} 
\label{dm}
\end{minipage}
\hspace{0.04\linewidth}
\begin{minipage}[b]{0.41\linewidth}
\caption{
The invariant mass distribution of dilepton production in NA50 
experiment~\protect\cite{na50}. 
}
\end{minipage}
\begin{minipage}[t]{80mm}
\epsfxsize=3.25in
\epsfbox{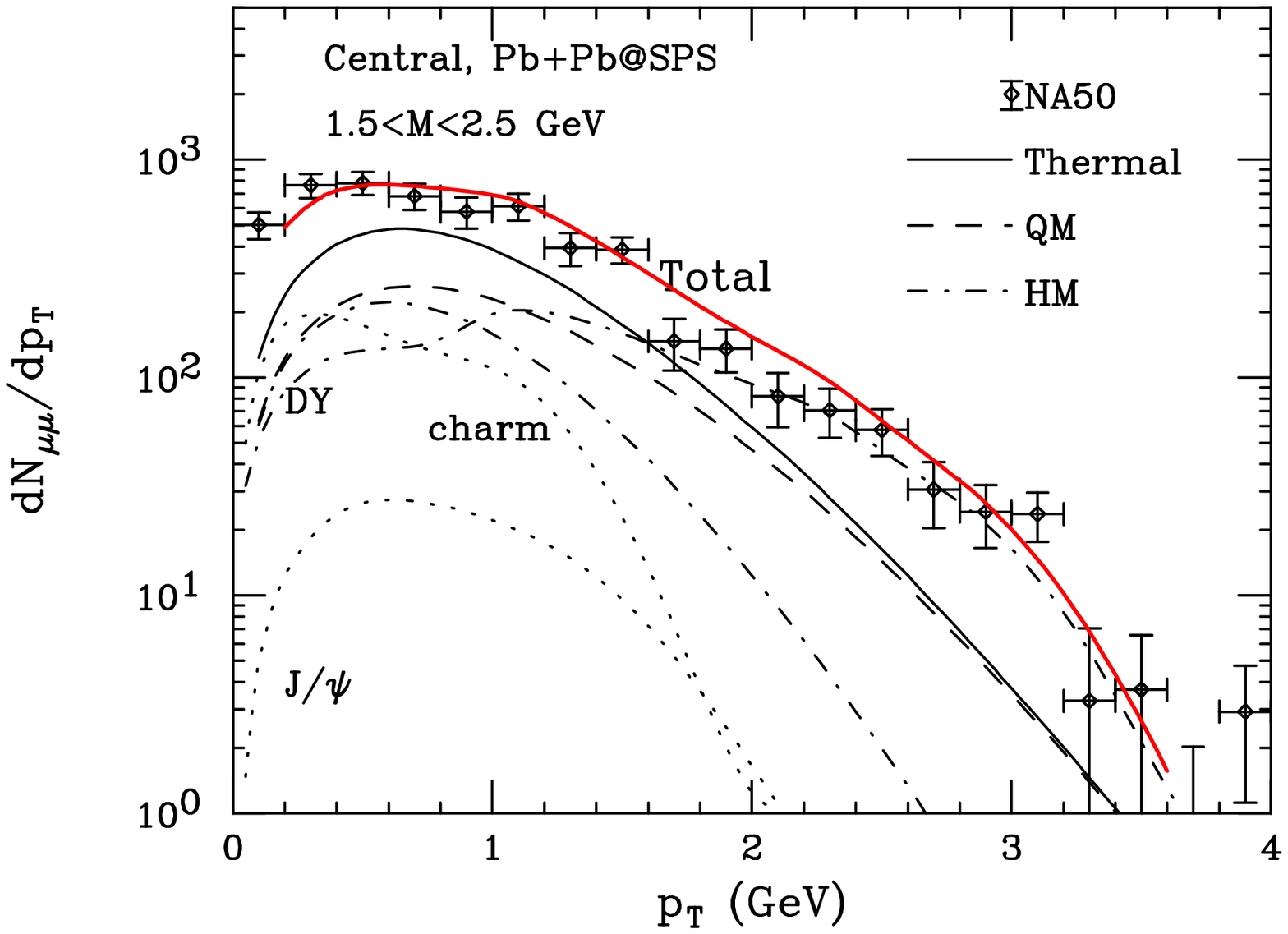}
\label{dpt}
\end{minipage}
\hspace{0.04\linewidth}
\begin{minipage}[b]{0.41\linewidth}
\caption{ The \protect$p_T$ distribution of dileptons produced
in central collision of lead nuclei.} 
\end{minipage}
\end{figure}
The plasma is assumed to consist of massless quarks (u, d, and s) 
and gluons with the number of flavours as
$\approx$ 2.5 to account for the mass of the strange quarks.
 We assume a rapid thermalization~\cite{kms} so that
the formation time is decided by the uncertainty relation and $\tau_0=1/3T_0$.
The energy density profile is assumed to be proportional to the wounded
nucleon distribution.
We further assume that the phase transition takes place at $T=$ 180 MeV and the
freeze-out takes place at 120 MeV.
 The rates for the hadronic matter have been
taken from Ref.~\cite{joe} and the 
contribution of the $A_1$ resonance is included according to the
suggestions of Xiong et al~\cite{li}. The relevant hydrodynamic equations are
solved using the procedure~\cite{hydro} discussed earlier and
an integration over history of evolution is performed~\cite{crs}.

In Fig.~1 we show our results for single photons. The dashed curve gives the
contribution of the quark-matter and the solid curve gives the
sum of the contributions of the quark matter and the hadronic matter.
The NLO pQCD estimate for prompt photons, with the inclusion
of intrinsic partonic momenta are also given~\cite{ktcr}.
We see that the thermal sources contribute about 50\% of the single
photons and that a very good description  of the data is obtained
when the thermal and the prompt sources are added.
Consequences of 
variation of some of the parameters can be seen in Ref.~\cite{hep}.
The fit to pion spectra from the WA98 experiment~\cite{wa98pi},
 kaon and proton spectra from the NA44 experiment~\cite{na44}, and the
$\Lambda+\Sigma$ spectra from the NA49 experiment~\cite{na49} are given in 
Fig.~2. 
The production of intermediate mass dileptons measured by the NA50
experiment is shown in Fig.~3. We see that the sum of the thermal
and Drell-Yan contributions provides a good description to the experimental
data. We add that we have used the procedure described in Ref.~\cite{rapp}
to simulate the detector acceptance. We also add that the thermal production
is quite identical to the enhanced production of charm decay estimated
by the NA50 group to `explain' this excess. The corresponding fit
to the $p_T$ spectrum is shown in Fig.~4. It should be noted that contrary
to the findings of Ref.~\cite{rapp} most of the radiations in the present
work comes from the quark matter itself. This difference is most likely due
to the rich equation of state along with a sophisticated evolution
mechanism for the plasma employed  in the present work Ref.~\cite{dil}.

In brief, we have shown that a single set of initial conditions,
which involve a quark gluon plasma in the initial state and envisage a
quark-hadron phase transition during the evolution, are
able to provide a consistent description to single photons, dileptons, and
hadrons produced in central $Pb+Pb$ collisions measured at the CERN SPS.
This we feel, provides a very strong support to the claim that a quark gluon
plasma is formed in these experiments.

We thank T. Awes and E. Scomparin for useful discussions.

\end{document}